\documentclass[a4paper,twocolumn,11pt,accepted=2020-08-31]{quantumarticle}
\pdfoutput=1
\usepackage[utf8]{inputenc}
\usepackage[english]{babel}
\usepackage[T1]{fontenc}
\usepackage{amsmath}
\usepackage{hyperref}
\usepackage{tikz}
\usepackage{lipsum}

\usepackage{amsthm, amssymb} 
\usepackage{graphicx}

\usepackage{color}

\usepackage[numbers,sort&compress]{natbib}

\usepackage{etoolbox}
\usepackage[outdir=./pix/]{epstopdf}
\usepackage{braket}

\newtoggle{localCompile}
\newtoggle{pasteBBLHere}

\usepackage{hyperref}

\begin{document}

\title{Automatic design of Hamiltonians}
 
\author{Kiryl Pakrouski$^{1}$}
\affiliation{$^{1}$Department of Physics, Princeton University, Princeton, NJ 08544, USA}
\orcid{0000-0003-0384-5909}

\begin{abstract}
We formulate an optimization problem of Hamiltonian design based on the variational principle. Given a variational ansatz for a Hamiltonian we construct a loss function to be minimised as a weighted sum of relevant Hamiltonian properties specifying thereby the search query. Using fractional quantum Hall effect as a test system we illustrate how the framework can be used to determine a generating Hamiltonian of a finite-size model wavefunction (Moore-Read Pfaffian and Read-Rezayi states), find optimal conditions for an experiment or "extrapolate" given wavefunctions in a certain universality class from smaller to larger system sizes. We also discuss how the search for approximate generating Hamiltonians may be used to find simpler and more realistic models implementing the given exotic phase of matter by experimentally accessible interaction terms.  
\end{abstract}


\maketitle

\section{Introduction}

In quantum physics one often starts by postulating a Hamiltonian of the system and studying its properties. Inventing a simple minimal Hamiltonian capable of capturing a physical phenomenon is an art and an important step towards design and control of quantum systems. 

Nature often operates with Hamiltonians that include a mix of such simple terms. 
Navigating this "Hamiltonian space" efficiently is crucial for designing a (numerical) experiment that best exhibits the physics we are after. In addition, because of the often non-trivial interplay of simple model terms, combining them we could as well find completely new phenomena as exemplified by the recent discovery of the many-body-localised phase~\cite{MBLGornyi2005,MBLBASKO2006,MBLtransitionHuse2010} found for the disorder, hopping and interaction terms mixed in an appropriate proportion. 
  
  A substantial progress~\cite{EDLectureNotes2008,QMCScience1986,DMRGWhitePRL92,ORUSTensorNetworks2014} on the central problem of solving a given quantum Hamiltonian numerically has been achieved in recent decades. Variational principle is widely used in order to minimise $\braket{\psi_{ansatz}(\{\gamma_i\})|H|\psi_{ansatz}(\{\gamma_i\})}$ yielding an approximation to the ground state $\ket{\psi_{ansatz}(\{\gamma_i\}^{optimal})}$ determined by the optimal values of the variational parameters $\{\gamma_i\}$. Various methods differ by the choice of the ansatz and minimisation strategy and can be purely numeric or utilise quantum hardware~\cite{quantumVariationalEigenvalueSolverNatComm} to evaluate the Hamiltonian expectation value~\footnote{\label{myfootnote} The way more general practical classical optimisation tasks can profit from using quantum hardware is exemplified by the Quantum Approximate Optimisation Algorithm~\cite{farhi2014quantumApprOptProbl}}. For many classes of systems we now have a tool capable of solving a Hamiltonian and deducing its relevant properties. In this situation one may wonder if we could systematically look for the Hamiltonian that optimises the said properties thus making the Hamiltonian dependent on variational parameters instead of the wavefunction.

 In this work we present a variational framework that automates the search for Hamiltonians that are best in some user-defined sense. Examples of the search queries may include:  "What is the most stable Hamiltonian with certain quasiparticle braiding statistics?" or "How do I combine the limited number of experimentally accessible interaction terms such that the many-body ground state is in the desired universality class?" The applications we will illustrate are finding parent Hamiltonians for model states, "extrapolating" the model states to larger system sizes and identifying the optimal conditions for an experiment.

One of these tasks, finding the generating (entanglement) Hamiltonian given an eigenstate attracted interest in literature recently~\cite{ClarkMethod2018,XiaoLiangMethod2016,ThomaleMethod2018,EntanglementGuidedSearch2019Prl,WeiCovReconstrEntHamPRB2019}. Given a reference wavefunction 
one chooses a set of hermitian operators $\hat{h}_i$ that form an operator basis in the considered Hilbert space or its subspace such that the Hamiltonian of interest can be expanded as $H = \sum_i \alpha_i \hat{h}_i$. Using the basis operators and the reference wavefunction a "covariance" matrix is formed whose 0-eigenvalue eigenvectors (if they exist!) specify the real coefficients $\alpha_i$ providing the expansion of a unique local hermitian operator for which the original wavefunction is an eigenvector. A more experimentally-relevant approach~\cite{PRL2019LearnLocHfromLocalMeas} constructs an analog of the "covariance" matrix based on the measurement of sufficiently many locally-supported operators in an eigenstate or a Gibbs state but doesn't need the access to the full wavefunction. In general, the possibility of recovering the full Hamiltonian from a single eigenstate is best understood for local and ergodic systems~\cite{GarrisonGroverETHvsHFromEigenstate,SwingleReconsStFromLocalData} (obeying some form of the eigenstate thermalisation hypothesis~\cite{deutsch1991quantum,srednicki1994chaos,rigol2008thermalization}).

Numerical optimization may be used to solve the inverse problem for the ground states of lattice Hamiltonians described by quantum field theories with emergent Lorentz invariance~\cite{EntanglementGuidedSearch2019Prl} or to find an effective Hamiltonian that best reproduces a given spectrum or entanglement spectrum~\cite{ConstructionOfHamiltoniansBySupervisedLearning}.
The problem of finding a density matrix that fits a given reference density matrix~\cite{KappenQBMArxiv2018,MelkoQBMPRX2018,WiebeQBM2017PRA} has been considered for lattice spin models and named Quantum Boltzman Machine. It may be used to approximate the Hamiltonian that generates a given density matrix.

Quantum device certification or inferring the Hamiltonian implemented in a particular experimental device is an adjacent field on the experimental side where the knowledge of the full wavefunction is usually not assumed. The input is rather the outcomes of the measurements performed on the device and one of the main challenges is the reduction of the number of measurements and computational resources required to restore the Hamiltonian. Progress in this direction has been reached using the compressive sensing techniques for generic Hamiltonians~\cite{compressiveSensingPRA2011} and by constraining a Hamiltonian ansatz with the outcomes of local measurements~\cite{PracticalCharacterizPRL2011}. Further efficiency improvement may be achieved by utilising the Bayessian experimental design ideas to choose the most "informative" measurements while the experiment is being made avoiding the need of storing previous results and delivering confidence regions along with the estimate of the most likely Hamiltonian parameters~\cite{GranadeOnlineHamiltonianLearning2012}. Bayessian inference has also been used to estimate the Hamiltonian based on the parameter dependence of certain observables~\cite{HukushimaBayesianInferenceOfHamiltonian}. Yet another approach to inferring an experimental Hamiltonian given a guess for its form uses a neural network-based protocol ~\cite{EvertHLeraningArxiv}.

Finding parent Hamiltonians based on the "covariance" matrix construction~\cite{ClarkMethod2018,XiaoLiangMethod2016,ThomaleMethod2018,PRL2019LearnLocHfromLocalMeas} is appealing as the least computationally demanding method. The main problem of this approach is stated in Ref.~\cite{ThomaleMethod2018}: "An obvious drawback is that the method guarantees that $\ket{\psi_0}$ is an exact or approximate eigenstate of $H$, but not that it is the ground state". Ref.~\cite{ThomaleMethod2018} further provides an example where the "covariance" matrix-based method finds a symmetry operator and not a Hamiltonian. In contrast, the method described in this work avoids both problems by design. When applied to find a parent Hamiltonian it ensures that the given reference state is approximated by the ground state of the Hamiltonian found.

Our method does not rely on building a "covariance" matrix or its analogues which allows us to generalise the way the Hamiltonian is parametrised. In fact, we only assume that the map from the space of real variational parameters into the space of Hermitian operators is continuous but do not require that the variational parameters are the basis expansion coefficients. This greatly broadens the range of possible applications of the method beyond finding parent Hamiltonians with one of such applications discussed in Sec.~\ref{sec:wk}.

In the context of searching for parent Hamiltonians the idea of using numerical optimisation of the basis expansion coefficients in cases when the "covariance" matrix approach doesn't yield an exact parent hamiltonian has been expressed in Ref.~\cite{ThomaleMethod2018}. The authors report however that the parameter updates, they used (based on the linear Newton scheme) did not yield a stable or robust method. In this work, we point out that the robustness of the optimisation is directly linked to how well-behaved the optimisation landscape is, the property that can be controlled by the loss function composition. The loss function combining multiple terms allows us to successfully apply optimisation for finding parent Hamiltonians. For the reference states we consider we are able to reproduce (see Sec.~\ref{sec:MRPf} and~\ref{sec:RR}) the exact and approximate generating Hamiltonians known in literature from analytical studies or manual parameter swipes.  

In the context of another application, the optimal experiment design, we further illustrate the importance of an organised update strategy. In particular, we show how even the non-linear gradient descent requires unpractical number of steps for convergence compared to the non-linear conjugate gradient descent (used in this work) even in a simple two-dimensional landscape considered in Sec.~\ref{sec:wk}. It is conceivable that the linear Newton scheme considered in Ref.~\cite{ThomaleMethod2018} would land in an irrelevant local minimum of a rough landscape generated by a single-target loss function when not accompanied by restarts or any other strategy for exploring nearby valleys.

For the task of parent Hamiltonian finding our approach is new in that it seeks to optimize a weighted combination of several measures as opposed to using only one of them and relaxes some physical assumptions such as emergent Lorentz invariance. Additional terms allow us to refine the search and look for the Hamiltonian that is not only in the same phase of matter as the reference state but also has other favourable properties such as energy gap or entanglement entropy. We can work with any continuous parametrisation and find a solution even when the set of available Hamiltonian terms is limited.

Specification of the terms entering the loss function and their weights determines the particular application. We note that while the presence of a term deriving from a certain reference wavefunction such as overlap is necessary for finding parent Hamiltonians no reference state is needed for many other applications. For example, one may be interested in "the most chaotic" Hamiltonian in a certain variational family. One might then consider the spectral average of nearby gaps ratio as a measure of chaos~\cite{levelSpacingvaluesPRL2013}, optimising which would require no reference wavefunction. It is the flexibility in the loss function design that leads to the generality of the method being presented. At the same time, as discussed in Sec.~\ref{sec:OptimisationProblem} the presence of multiple terms in the loss function improves the convexity properties of the optimisation landscape and renders the optimisation approach more practical and robust.
 
We describe our framework in Section~\ref{sec:Method}, briefly introduce fractional quantum Hall effect, our test system, in Section~\ref{sec:FQHE} and present some of the possible applications in Section~\ref{sec:Applications}.

\section{Method\label{sec:Method}}

\subsection{Hamiltonian parametrisation}

We assume the existence of an ansatz for the Hamiltonian that continuously maps a parameter vector $\{\gamma_m\}$ from a convex subregion of $\mathbb{R}_M$ into a valid Hamiltonian operator.

One way to achieve this is by linearly combining $M$ "basis" operators $\{\hat{O}^b_m\}$ into a sum with coefficients given by $M$ real parameters $\gamma_m$:

\begin{equation}
\label{eq:Hexpansion}
H = \sum_{m=1}^{M} \gamma_m \hat{O}^b_m.
\end{equation}

This form is useful if we would like to consider Hamiltonians made of certain type of terms (say, only translationally invariant Hamiltonians with nearest-neighbour couplings). In the experimental context $\{\hat{O}^b_m\}$ may be the set of all experimentally accessible interaction terms.

On the other hand as discussed in Sec.~\ref{sec:wk} the Hamiltonian parametrisation may be given by a theoretical model that constructs an effective Hamiltonian given the values of the variational parameters $\{\gamma_m\}$. In any case we assume some pre-defined general structure of the Hamiltonian such that the number of variational parameters only grows polynomially with the system size, not exponentially. 

\subsection{Loss function}

The ultimate goal is to search the variational space of Hamiltonians and find the optimal point $\{\gamma^o_m\}$ minimising a loss function $LF$ made of a weighted sum of several observables. We define below the observables used in this work whereas any real, single-valued, piece-wise continuous functions of the variational parameters may be used in general.

\textbf{Relative Energy variance} is defined for a given reference wavefunction $\ket{\psi_{r}}$ and a Hamiltonian $H_O$ as
\begin{equation}
(\sigma^{r}_E)^2 = \frac{ \braket{\psi_{r}|H_O^2|\psi_{r}} - \braket{\psi_{r}|H_O|\psi_{r}}^2}{ \braket{\psi_{r}|H_O|\psi_{r}}^2}.
\end{equation}
In case $\ket{\psi_{r}}$ is an eigenstate of $H_O$ this quantity is zero. Otherwise its magnitude can be used as a measure of how close is $\ket{\psi_{r}}$ to being an eigenstate of $H_O$. We will quote $(\sigma^{r}_E)^2$ in our results however in the optimisation we will use $\sigma_E^2 = \braket{\psi_{r}|H_O^2|\psi_{r}} - \braket{\psi_{r}|H_O|\psi_{r}}^2$,  which ensures that the minimisation of relative energy variance is not achieved by an increase in the ground state energy.

We use two measures of similarity between two wavefunctions. One is simply the absolute value of their \textbf{overlap} or scalar product: $|\braket{\psi|\psi_{r}}|$. Another measure can be defined if we view the two wavefunctions as probability distributions in the Hilbert space. The distance between such distributions can be defined as their relative entropy also known as Kullback-Leibler (\textbf{KL}) \textbf{divergence}: 
\begin{equation}
D_{KL} = \sum_i \alpha_i^2 log( \frac{\alpha_i^2}{\beta_i^2} ),
\end{equation}
where $\alpha_i$ and $\beta_i$ are the expansion coefficients of the two wavefunctions in a basis $\ket{b_i}$ spanning the Hilbert space: $\ket{\psi_{r}}=\sum_i \alpha_i\ket{b_i}$ and $\ket{\psi}=\sum_i \beta_i\ket{b_i}$. In case two wavefunctions are identical their overlap is one and the KL-divergence is 0. The KL-divergence only depends on the absolute values of the wavefunction coefficients and not their signs thus, its zero value is a necessary but not sufficient condition for two wavefunctions to be identical.

Further observables that will be used include: ground state energy $E_0$, energy gap $\Delta_N$ for a particular system size N, energy gap extrapolated to the thermodynamic limit $\Delta_{extr}$, symmetry generator quantum number evaluated in the ground state.

In general, for any observable $T(\{\gamma\})$ a term proportional to $(T(\{\gamma\})-T_{ref})^2$ may be added to the loss function to select the solutions where the value of the observable is close to the desired value $T_{ref}$.

The loss function $LF$ will be a weighted sum of the chosen observables evaluated for a particular finite system size $N$: $t^N_k$ or system-size independent properties $T_s$
\begin{equation}
\label{eq:tf}
LF=\sum_k w_k \sum_N G_N t^N_k + \sum_s w_s T_s,
\end{equation}
where $G_N$ are constant coefficients used to prioritise certain system sizes if needed (in most cases we set $G_N$ proportional to the Hilbert space dimension) and the weights $w_k$ and $w_s$ are hyper-parameters defining the importance of an observable.

A regularisation term proportional to
\begin{equation}
\label{eq:regulTerm}
\sum_{m=1}^M |\gamma_m - \gamma^{ref}_m|
\end{equation}
may be added to the loss function to look for a solution in the vicinity of a certain reference point in the parameter space $\{\gamma^{ref}_m\}$.

In case one has access to the full finite-temperature density matrix of the system the difference between reference and any other density matrices may be measured by their relative entropy~\cite{RelEntropyReviewVadral2002,WiebeQBM2017PRA,KappenQBMArxiv2018}.

We can define the desired search region in the parameter space by adding a term quickly growing outside of it to the loss function. This may depend on the application and is especially useful when $\{\gamma_m\}$ are experimentally tuneable parameters for which a natural range of allowed values exists.

To choose the weights for all the terms in the loss function one may use the following strategy. First, we can construct the loss function where all the terms have approximately the same importance. Let's assume that a loss function term $T_s$ has the optimal value $T^{opt}_s$ near the solution being sought. Obviously, we don't know this value exactly but for our purposes knowing the order of magnitude suffices. For example, one may expect the wavefunction overlap to be around unity and the energy variance to be on the order of $10^{-5}$ for a reasonable solution. To have all the terms be of comparable importance we may start by choosing the importance weight $w_s=1/T^{opt}_s$. From here, further adjustment of the weights may be made depending on our final goal by introducing larger than one factors for the measures of particular importance. In this work we observed that the solutions obtained do not strongly qualitatively depend on the weights choice. We do however list in the supplementary the exact values used to simplify reproducing the results.

In principle, optimisation over the weights defining the loss function may be made an external loop around the algorithm described here. This external optimisation may be targeted to fine-tune the result according to the trade off that may arise between the important terms of the loss function. The details of this trade off and the strategy of handling it will be highly dependent on the particular application and the desired result. The optimisation over the weights would also introduce a factor in the computational cost proportional to the number of weights. For these reasons we refrain from implementing the external weight optimisation in this proof-of-principle work and defer it to the future application-oriented studies.

In this work we will use exact diagonalisation to evaluate the observables of a given Hamiltonian. Any other numerical tool suitable for a particular application may be used instead. 

\subsection{Optimisation problem}
\label{sec:OptimisationProblem}

We formulate the optimisation problem as follows: find the set of $\{\gamma_m\}$ from a convex subregion of $\mathbb{R}_M$ that minimises the loss function defined in Eq.~\ref{eq:tf}.

Ideally, the loss function should be designed such that the optimisation problem has a unique solution and is convex.
A prominent feature of non-convex problems is that there may exist multiple solutions to the problem and finding the global minimum may be very hard. Although it may appear problematic from the abstract mathematical point of view also in this case we can obtain useful results by following several simple strategies.

First, a non-convex function may be convex on a sub-domain. A simple example is a cosine which has many minima on $[-\infty,+\infty]$ but only one on $[0,2\pi]$. Adding a regularisation term (Eq.~\ref{eq:regulTerm}) to the loss function that penalizes deviations from a certain reference Hamiltonian is a way to profit from this property. In the absence of a meaningful reference Hamiltonian a trivial zero may be used rendering the search to be the search for the simplest, minimal model. In the experimental context there is often a well defined preferred region of experimental parameters. Using a point in that region as a reference we search for the minimal deformation of the reference interaction that optimises the properties of interest.

All the symmetries of the loss function w.r.t. the variational parameters should be explicitly broken by adding a corresponding term. Constructing the loss function out of multiple physical variables may be seen as another way of increasing the chance of a unique solution.

Various general techniques have been developed in the field of non-convex optimisation. One can start the optimisation from a several different points and pick the minimum solution should the converged results differ. Evaluating the Hessian matrix may be used to escape a saddle point. Adding some "temperature" over a few parameter updates may help escape from a deep local minimum.

Finally, for some applications finding a solution may be of greater value than ensuring it is unique.  

\subsection{Optimisation algorithm}

We use the non-linear conjugate gradient descent~\cite{CGFletcherReeves1964,ConjGradAndNonConvexBook1984} algorithm (CGD) with restarts~\footnote{we are willing to provide the implementation code upon request} which can be substituted with other optimisation schemes as appropriate for the application.

Conjugate gradient descent is an iterative first order method where the iterations start with an initial guess for the solution $\{\gamma^0_m\}$, which can be generated randomly. Next, the gradient of the loss function w.r.t. all of the $\{\gamma_m\}$ is calculated. Where possible the partial derivatives of the loss function should be calculated analytically. The gradient can also be estimated numerically by evaluating the loss function at a small displacement from the initial point $\{\gamma^0_m\}$. More explicitly, we may approximate the $m$'s component of the gradient by the finite difference $\frac{LF(\vec{\gamma^0}) - LF(\vec{\gamma}^0 + \hat{i}_m \delta)  }{ \delta}$, where $\hat{i}_m$ is the unit vector in the $m$s direction and $\delta$ is an increment, small enough to provide a good approximation for the derivative but large enough to avoid the numerical precision problems. In the worst (latter) case the computational effort grows linearly with the dimension of the parameter space $M$. 
In the applications discussed in this work the gradient was approximated using the above finite difference and $\delta$ was chosen to be $2*10^{-8}$. 

Once the gradient direction is calculated a step is made in the opposite direction. The size of the step is chosen s.t. it minimises the loss function the most, which we implement using golden section line search~\cite{goldenSectionLineSearch} combined with evaluating the loss function on four additional points along the gradient. The points are chosen to lie a distance [$10^{-9}$ $10^{-7}$ $10^{-5}$ 0.1 0.99]$a$, with $a$ the distance between the current point in the parameter space and the boundary of the reasonable region to be considered. This "non-linearity" is intended to allow the updates to cut through possible local bumps on the way to the global minimum. Further updates move in the "conjugate" direction which admixes the history of previous updates to the direction of the current gradient (see Sec.~\ref{sec:wk} for an illustration).

Special treatment may also be needed at the points of the loss function discontinuity that may occur if the loss function contains, for example, terms proportional to the discrete quantum numbers. The jumps of the loss function may be signaled by the "explosion" of the gradient magnitude and verified by comparing the values of the discrete loss function terms on the two sides of the possible discontinuity. A heuristic step in the parameter space may then be chosen to cross the discontinuity and resume the conjugate gradient iterations on the side corresponding to the smallest value of the loss function.

We used the Hestenes-Stiefel~\cite{CGHestenesStiefel1952} and Polak-Ribeire schemes~\cite{polakRibeireScheme1969} for calculating the conjugate direction and reset the direction to pure steepest descent every $M$ steps.

\section{Test system: Fractional quantum Hall effect \label{sec:FQHE}}

In this work we illustrate a few applications of the described framework on the example of the fractional quantum Hall states and Hamiltonians and we briefly introduce them here.

The fractional quantum Hall effect~\cite{FQHEGirvinBook,CompositeFermionsBook} (FQHE) is due to strongly-correlated many-electron states formed in 2D electron gas subject to a perpendicular magnetic field. In strong field the resulting Landau levels are well separated and the non-trivial physics can be approximately described by the Coulomb interaction within the topmost partially filled Landau level. Thus, ignoring dimensional constants the Hamiltonian in real space may be written simply as 
\begin{equation}
H = \sum_{i<j} \frac{1}{|r_i-r_j|}.
\end{equation}
To avoid having boundaries that can lead to severe finite-size effects one often works on the spherical shell enclosing a magnetic monopole of strength $2S$~\cite{HaldanePPs1983}, where the solution to a one-particle Schroedinger equation is given by the generalised spherical harmonics $Y_{Slm}$~\cite{CompositeFermionsBook}. Angular momentum projection $m \in [-l,l]$ is the only free quantum number given fixed system size and Landau level index. Thus a many-body fermionic state can be thought of as a one-dimensional chain of zeros and ones corresponding to the presence or absence of an electron in a certain "orbital". Coulomb interaction corresponds to a (momentum conserving) pairwise interaction between all the filled orbitals. The radius of the sphere will grow with system size as $R\sim\sqrt{N_e}$, with $N_e$ - number of electrons.

A very useful way of parametrising any centre of mass momentum-preserving interaction in this setting is given by the Haldane pseudopotentials $V_m$~\cite{HaldanePPs1983,PPsManyBodySimon2007}
\begin{equation}
\label{eq:defPPs}
H = \sum_{m=1}^{2l} V_m \hat{P}^{(2b)}_m,
\end{equation}
where the operators $\hat{P}^{(2b)}_m$ project each pair of electrons into a state with a fixed total angular momentum $m$. The set of numbers $V_m$ thus defines an interaction and the pseudopotentials corresponding to the Coulomb interaction are well known~\cite{CompositeFermionsBook}. For spin-polarized systems, we will be concerned with in this work, only odd values of $m$ in the sum~\ref{eq:defPPs} are allowed due to fermionic statistics. In addition to the 2-body one may consider 3-body terms that project the triples of electrons into a fixed total angular momentum state. We provide the matrices corresponding to the operators $\hat{P}^{(2b)}_m$ in an explicit text format as supplementary files. The reference wavefunctions used in this work are provided as well.

One may use the pseudopotentials $V_m$ as the variational parameters $\gamma_i$ and the projectors $\hat{P}^{(2b)}_i$ as a set of the allowed operators $\hat{O}^b$.

To stabilise the optimisation we would like to exclude the trivial transformations of the Hamiltonian of adding or multiplying it by an overall constant. Such transformations do not change the nature of the many-body ground state. They may be avoided by fixing one of the pseudopotentials to 1 or by fixing the sum of all the absolute values of all the pseudopotentials.

The valid many-body states in the spherical geometry, used throughout this work, are eigenvectors of the total angular momentum $\hat{L}$ and the requirement that the state is spatially uniform corresponds to the condition $L=0$. A term $w_L L^2$ may be added to the loss function to enforce this.  

\section{Applications \label{sec:Applications}}

\subsection{Finding (approximate) generating Hamiltonians from an eigenstate \label{sec:findingGenH} / finding simpler, more realistic ways to implement a given phase of matter}

Given a reference wavefunction $\ket{\psi_{r}}$ we would like to find the best approximation to its generating Hamiltonian expanded as a linear combination of a given set of hermitian operators $\{\hat{O}^b\}$: $H=\sum_m \gamma_m \hat{O}^b_m$. The generating Hamiltonian must approximate the reference wavefunction with its ground state $\ket{\psi_{o}}$. Besides finding the actual parent Hamiltonians this formulation can also be used to find simpler realistic Hamiltonians that implement the same phase of matter as the true parent Hamiltonian but only involve the chosen set of terms $\{\hat{O}^b\}$ that can be implemented in a particular experimental setting or preferred on other grounds.

We define an approximation to the generating Hamiltonian as the Hamiltonian that minimizes the loss function from Eq.~\ref{eq:tf}, where the largest weights are given to the overlap $\braket{\psi_o|\psi_{r}}$, KL-divergence and energy variance. Further terms suitable for a particular case may be used to measure the proximity of the solution to the desired phase. If the solution is not unique (may be the case for non-local Hamiltonians) we take the solution closest to the given reference interaction as an answer. 

By construction, the Hamiltonian found approximates the desired model wavefunction with its ground state as opposed to a state in the middle of the spectrum; there will always be at least one solution and we do not risk to find a symmetry operator instead of the Hamiltonian. These are the major distinctions between our framework and the covariance-matrix-based methods described in literature recently~\cite{ClarkMethod2018,XiaoLiangMethod2016,ThomaleMethod2018}. On the other hand we make no assumptions about the relation between the parent Hamiltonian and the reduced density matrix as opposed to Ref.~\cite{EntanglementGuidedSearch2019Prl}. The generality of the method comes with potentially higher computational cost and it may be reasonable to attempt a covariance-matrix method~\cite{ClarkMethod2018,XiaoLiangMethod2016,ThomaleMethod2018} first or use its result as a starting point for the iterations in our approach although we didn't do so.

\subsubsection{Moore-Read Pfaffian state}
\label{sec:MRPf}
 \begin{figure}[h!]
                \centering
                \iftoggle{localCompile}{
                \includegraphics[width=9cm]{pix/pf2and3bConvergence12el}
                }{
                \includegraphics[width=9cm]{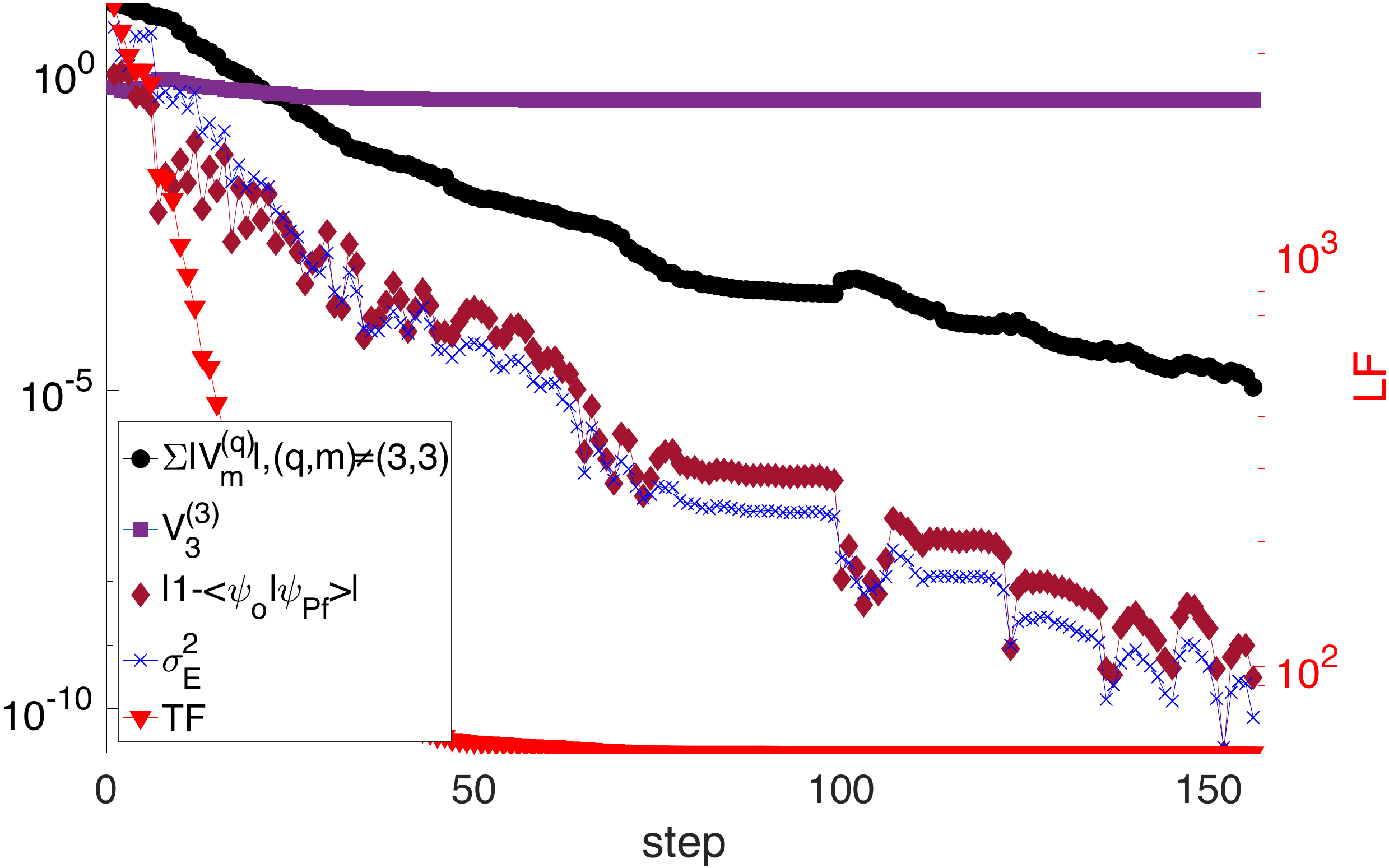}
                }
                \caption{Evolution of the Hamiltonian parameters in the search of the MR Pfaffian generating Hamiltonian with the CGD algorithm steps (horizontal axis). Hilbert space dimension is 16'660 while the parameter space is 27-dimensional. The loss function (red triangles) corresponds to the right vertical axis whereas all other data corresponds to the left vertical axis. Purple squares show the value of the $V_3^{(3)}$ pseudopotential, the only term in the Hamiltonian needed to generate the reference wavefunction. Black dots show the total value of all other pseudopotentials. The error in the overlap and the energy variance are shown with dark red diamonds and blue crosses respectively.}
                \label{fig:Pf3bHistory}
\end{figure} 

The Moore-Read Pfaffian state~\cite{MRPfaffian1991} (MR Pfaffian) is one of the candidates to explain the $\nu=5/2$ FQHE~\cite{HaldaneRezayiPfAt5half,fivehalfPRXPakrouski2015,RezayiV9}. It describes a gapped topological strongly-correlated phase of electrons with quasi-particle excitations governed by non-Abelian braiding statistics.  The MR Pfaffian state can be exactly generated as a ground state of the 3-body interaction where only a single 3-body pseudopotential corresponding to the closest approach of 3 electrons $V^{(3b)}_{M=3}$ is non-zero~\cite{GreiterHPfPRL1991,GreiterPfHNuclB1992} (originally verified with the help of the method described in Ref.~\cite{ThomaleMethod2018}). In this section we test if we can establish this fact automatically by means of optimisation.

Let the 12-electron MR Pfaffian wavefunction on the sphere be the reference state (in this example we will use a single system size for training). For this system $m \in [-10.5,10.5]$ and the Hilbert space dimension is 16'660.
We consider the set $\{\hat{O}^b\}$ that includes all 2-body and eleven lowest 3-body pseudopotentials: $V^{(3-body)}_{M=3}$ through $V^{(3-body)}_{M=14}$. Even-$M$ 2-body and $M=0,1,2,4$ 3-body pseudopotentials are excluded by fermionic statistics for spin-polarized systems. The 3-body pseudopotentials corresponding to $M=9,11,12,13,14$ are multi-valued and we parametrise each of them with 2 diagonal entries of a $2\times2$ matrix. Thus in total we have 11 (2-body) and 16 (3-body) free parameters and the dimension of the parameter space in which we search for the generating Hamiltonian is 27. The loss function $LF$ was mainly composed of the overlap, KL-divergence, regularisation ($\gamma^{ref}_i=0$) and energy variance (see Table 1
in Supplementary Materials). The pseudopotentials are initialised with random numbers uniformly distributed between 0 and 1 which means we input no prior knowledge of the expected form of the resulting Hamiltonian.

The convergence towards the correct solution is shown in Fig.~\ref{fig:Pf3bHistory}. Already after 45 steps all the "unnecessary" pseudopotentials are below $10^{-2}$ while the value of $V^{(3b)}_{M=3}=0.3708$, the overlap between the ground state and reference wavefunction is 0.999917 and the energy variance is 4.31*$10^{-5}$. We terminate the iterations after 156 steps when the overlap of 0.9999999997 and the energy variance of 7.3*$10^{-11}$ are reached.

We have demonstrated that in cases when the chosen set $\{\hat{O}^b\}$ includes all the operators necessary for exactly generating a wavefunction we are able to find the exact parent Hamiltonian. In the remainder of this Section we consider a \textbf{more realistic problem formulation}: find the Hamiltonian that generates the best approximation to the desired wavefunction given that the \textbf{accessible operator basis $\{\hat{O}^b\}$ does not include} all the terms needed to generate that wavefunction exactly. In this formulation the method may be applied to find simpler, experimentally relevant Hamiltonians generating the ground state in a given phase of matter.

To illustrate this scenario we ask what is the minimal \textbf{2-body} Hamiltonian that best approximates the MR Pfaffian with its ground state. The loss function includes contributions for system sizes between 6 and $N^{max}_{e~tr}$ electrons (with $N^{max}_{e~tr}$ between 10 and 16). To fix the energy scale we fix $V_1=1$. The optimisation results are summarised in Table~\ref{tab:genPpsPfMin}. We find that all the pseudopotentials $V_m$ converge to near-zero values for $m>3$. The value of the $V_3$ pseudopotential converges to values between 2.5 and 3 depending on $N^{max}_{e~tr}$.

The possibility of approximately generating MR Pfaffian using 2-body interaction was discovered earlier upon particle-hole symmetrising the contact 3-body interaction~\cite{PetersonH2ForPf2008} which cancels all 3-body terms exactly and also discussed in~\cite{JainPRB2017Pf2bH}. Here we arrive at this result by applying a generic optimisation procedure suitable for arbitrary Hamiltonians. Recently, it was shown~\cite{PetersonScarola2019} that the pseudopotential ratio $V_1/V_3$, resulting from the particle-hole symmetrisation~\cite{PetersonH2ForPf2008}, extrapolates to 3 with increasing the system size.  Our results are in agreement with prior literature whereas the overlaps we find are slightly higher than those in Ref.~\cite{PetersonScarola2019} (see Table I).  

\begin{table}
\caption{\label{tab:genPpsPfMin} Minimal two-body Hamiltonians found to produce closest approximation to the finite-size MR Pfaffian ($\ket{\psi_r}$) state. Here and in all the tables: 1. Data is grouped in columns according to the value of $N^{max}_{e~tr}$ ($N^{max}_{h~tr}$) - the number of electrons (holes) in the largest system used to learn the generating Hamiltonian 2. $\ket{\psi_o}$ is the ground state of the learned optimal Hamiltonian (may be computed for system sizes $N_e>N^{max}_{e~tr}$) 3. $D_{Hilbert}$ is the Hilbert space dimension of the largest system used to learn the Hamiltonian 4. $M$ is the number of variational parameters (see Eqs.~\ref{eq:defPPs} and~\ref{eq:Hexpansion}). }
\begin{center}
\resizebox{\columnwidth}{!}{%
\begin{tabular}{|c|c|c|c|c|}
\hline
$N^{max}_{e~tr}$ 						& 10			& 12			& 14			& 16 \\
\hline
$D_{Hilbert}$							& 1'514 		&  16'660		& 194'668   	& 2'374'753\\
\hline
$M$									& 9 			&  11			& 13   		& 15\\
\hline
$\braket{\psi_{o}|\psi_{r}}(N^{max}_{e~tr})$ 	& 0.9979038 	& 0.9901101	& 0.9836215	&  0.9724166 \\
\hline
$\braket{\psi_{o}|\psi_{r}}(N_e=18)$ 			& 0.9499857  	& 0.9536377	& 0.9553969	&  0.9601343 \\
\hline
$(\sigma^{r}_E)^2(N^{max}_{e~tr})$		& 8.011e-05	& 0.0001742 	& 0.0001750  	&  0.0001907 \\
\hline	
$V_1$:$V_3$, $V_3=1$					& 2.6514122	& 2.7448621	& 2.6533339  	&  2.6819474 \\
\hline	
$\sum_{m\ne1,3} |V_m|$					& 0.0091164	& 0.0000087	& 0.026952  	&  0.0550644 \\
\hline
\end{tabular}
}
\end{center}
\end{table}

\subsubsection{Read-Rezayi state}
\label{sec:RR}
The Read-Rezayi state~\cite{RRstate1999} (RR) is a possible description~\cite{ReadRezayi2009,RRDShengHaldaneDMRGPRL2015,Pakrouski201612fifth} of the $\nu=13/5$ FQHE (its particle-hole conjugate (cRR) - of the $\nu=12/5$ FQHE). It is a gapped topological phase with quasiparticle excitations governed by non-Abelian braiding statistics complex enough to implement universal quantum computing~\cite{NayakReviewQuantComp2008}. The RR state can be exactly generated as the ground state of the 4-body Hamiltonian with only contact 4-body interaction turned on. Here we again attempt to approximate the generating Hamiltonian of the RR state with much simpler interaction only involving 2-body terms. The systems with the number of holes between 4 and $N^{max}_{h~tr}$ (where $N^{max}_{h~tr}$ is from 8 to 12, see Tables~\ref{tab:overlapsRRMinimal} and~\ref{tab:overlapsRRCoulomb}) will be used in optimisation.

While looking for the minimal generating model we find that all $V_m$ for $m>7$ are irrelevant and converge to zero. The results for the optimal values of the remaining 3 pseudopotentials are presented in Table~\ref{tab:overlapsRRMinimal}.

The possibility to produce the approximation to the RR state as a ground state of a two-body Hamiltonian has been discussed in literature recently~\cite{Wojs2019RR2body,Wojs2018ParamScan2bodyForRR}. In ref.~\cite{Wojs2018ParamScan2bodyForRR} an extensive parameter scan of the 3 lowest pseudopotentials ($V_1,V_3,V_5$) was performed and in Ref.~\cite{Wojs2019RR2body} an analytic mean-field calculation. It was found that an approximation to the RR state can be generated using the lowest 3 pseudopotentials such that $V_1:V_3:V_5=6:3:1$ in the thermodynamic limit. The values we find are in agreement which is another test of our procedure.

Same as for MR Pfaffian, with increasing system size the optimal pseudopotentials gradually change with the changing curvature of the spherical shell on which the calculation is performed. For this reason, using the Hamiltonian learned on a small system, say $N^{max}_{e~tr}=8$ produces the ground state at larger sizes with high but not ideal overlap with the model wavefunction ($\braket{\psi_{o}|\psi_{r}}(14)$=0.8432350429). This geometrical finite-size effect is specific to the FQHE calculations in spherical geometry and should be absent in other setups.

\begin{table}
\caption{\label{tab:overlapsRRMinimal} Minimal two-body generating Hamiltonian for the Read Rezayi state ($\ket{\psi_r}$). System size is indicated by the number of holes $N_h$.}
\begin{center}
\resizebox{\columnwidth}{!}{%
\begin{tabular}{|c|c|c|c|c|}
\hline
$N^{max}_{h~tr}$						& 8					& 10					& 12 \\
\hline
$D_{Hilbert}$							& 12'346 	 			& 246'448   			& 5'216'252 \\
\hline
$M$									& 11					&  14					& 16   \\
\hline
$\braket{\psi_{o}|\psi_{r}}(N^{max}_{h~tr})$ 	&  0.9902541126		& 0.9593283571		& 0.970100531\\
\hline
$\braket{\psi_{o}|\psi_{r}}(14)$				&  0.8432350429		& 0.8423694609		& 0.899087368 \\
\hline
$(\sigma^{r}_E)^2(N^{max}_{h~tr})$			& 1.2191662e-05		& 2.244818e-05 		& 2.32343e-05 \\
\hline
$V_1$:$V_3$, $V_5=1$					& 5.9367:3.1327		& 6.0150:3.1724		&  5.7773:2.898\\
\hline	
$\sum_{m\ne1,3,5} |V_m|$				& 0.013350502			& 0.0000102429		&  0.009911585   \\
\hline
\end{tabular}
}
\end{center}
\end{table}

\subsection{Extrapolating ground states to higher system sizes}
Another application comes as a natural extension and test for finding the generating Hamiltonians. In general, the Hamiltonian we find is approximate.

If we used system sizes of up to $N^{max}_{e~tr}$ electrons for training than a good-approximation Hamiltonian would produce the ground states in the same phase of matter also for higher system sizes.

To quantify this definition we calculate the overlap between the true model wavefunction ($\ket{\psi_{r}}$) and the ground state of the learned optimal Hamiltonian ($\ket{\psi_{o}}$) at the system sizes beyond those used in training ($N_{e~test}>N^{max}_{e~tr}$). The Hamiltonian giving high overlaps has high "predictive power" and thus the optimisation scheme could be used to "extrapolate" the ground states in certain phase of matter to higher system sizes.

This approach is useful for the cases when the algorithm for obtaining the model wavefunction scales significantly faster with the system size then exact diagonalisation (particle-hole symmetric Pfaffian wavefunction~\cite{SonPRX2015} is an example). The direct construction of a wavefunction becomes then impossible beyond very small systems but one can obtain the wavefunction in the same phase for larger sizes by diagonalising the Hamiltonian learned on the directly constructed examples.

Here we investigate how successful this program may be in the (more general) case when the Hamiltonian ansatz does not include the terms that are needed to generate the desired wavefunction exactly. Throughout this section we use the 2nd Landau level Coulomb interaction as the reference and thus are finding the minimal deformation of the Coulomb interaction that reproduces the model wavefunctions.

\begin{table}
\caption{\label{tab:overlapsPfCoulomb} 
MR Pfaffian ($\ket{\psi_{r}}$) overlaps and energy variances computed for the optimal two-body Hamiltonian found as a minimal deformation of Coulomb interaction (the pseudopotentials defining $H_O$ can be found in Table VII
in the Supplementary Materials
). Here and in Tab.~\ref{tab:overlapsRRCoulomb} rows correspond to the system size where the learned model was tested. The results for the test sizes larger than the ones used to learn $H_O$ are below the horizontal separating line in each of the columns. All the energy variances ($\sigma^{r~2}_E$) are given in units of $10^{-7}$.}
\begin{center}
\resizebox{\columnwidth}{!}{%
\begin{tabular}{|c|c|c|c|c|c|c|c|c|}
\hline
$N^{max}_{e~tr}$							& \multicolumn{2}{c|}{10	 } 			& \multicolumn{2}{c|}{12	}		& \multicolumn{2}{c|}{14	}			& \multicolumn{2}{c|}{16	}\\

\hline
$N^{test}_{e}$								& $\braket{\psi_{o}|\psi_{r}}$& $\sigma^{r~2}_E$  		& $\braket{\psi_{o}|\psi_{r}}$&$\sigma^{r~2}_E$				& $\braket{\psi_{o}|\psi_{r}}$ & $\sigma^{r~2}_E$			& $\braket{\psi_{o}|\psi_{r}}$&$\sigma^{r~2}_E$ \\
\hline
6								& 0.98252& 18 						& 0.81519 & 21						& 0.88187 & 18					& 0.46987 & 18 \\
8								& 0.99097& 13 						& 0.93817 & 28 					& 0.94525 & 25 					& 0.9314 & 19 \\
10								& 0.99923& 3.1						& 0.99775 & 1.7 					& 0.99859 & 2.9					& 0.99765 & 2.6\\ \cline{2-3}
12								& 0.99362& 13						& 0.99493 & 3.5					& 0.98888 & 4.2 					& 0.99 & 3.1 \\ \cline{4-5}
14								& 0.94816&17						& 0.70441 & 11						& 0.98838 & 4.2					& 0.98734 & 3\\ \cline{6-7}
16								& 0.96209&15						& 0.81230 & 16						& 0.97892 & 4.3					& 0.98184 & 2.7\\ \cline{8-9}
18								& 0.95795&12						& 0.83458 & 10						& 0.94592 & 4.7					& 0.95924 & 3.3\\

\hline
\end{tabular}
}
\end{center}
\end{table}

The loss function is mainly designed to optimise the overlaps and the energy variances of the "training" system sizes (6 to $Ne^{max}_{tr}$ for MR Pfaffian). There is an additional importance factor on the order of 1000 assigned to the largest training system $N^{max}_{e~tr}$. As a result the optimisation is mainly looking to improve the overlap and energy variance for the largest training system. 

We would like to allow fluctuations for most of the pseudopotentials here. This is in contrast to searching for the minimal model (see Tables~\ref{tab:overlapsRRMinimal} and \ref{tab:genPpsPfMin}) where we would like the unimportant pseudopotentials to be almost exactly zero and only few important ones stand out. This means that the effective radius of the neighbourhood around the reference solution we would like to search is much larger here as compared to the minimal model search. This allowed neighbourhood radius is controlled by the inverse of the regularisation term weight in the loss function and it thus has to be much smaller in this case. The particular weights used for the calculations can be found in the supplementary materials.

We expect that the predictive power of the model should increase with the maximum size of the wavefunctions used in the learning phase due to the decreasing (quantum and geometric) finite-size effects which should make larger systems more representative of the thermodynamic limit. This is the trend we observe for both MR Pfaffian (Table~\ref{tab:overlapsPfCoulomb}) and RR (Table~\ref{tab:overlapsRRCoulomb}) states:  
the overlap and the energy variance improve with increasing the maximum training system size. Finite-size effects may disturb this tendency for small system sizes as can be seen on the example of the $N^{max}_{e~tr}=10$ MR Pfaffian state.

The Hamiltonian we learn can only reproduce the correlations present in the training systems. Should all the training systems be too small to fully accommodate a certain phase, the learned Hamiltonian will fail to approximate the ground states at the higher system sizes where all the correlations are fully manifested. This is the situation we observe for the RR state (see the first column of Table~\ref{tab:overlapsRRCoulomb} ): if only RR wavefunctions for 4,6 and 8 holes ($N_{\phi}$=12,17 and 22) are used the learned optimal Hamiltonian can only poorly approximate the RR state at higher sizes. Thus, if the extrapolation to higher system sizes is the goal one must train on the system sizes beyond the characteristic correlation length of the phase.


\begin{table}
\caption{\label{tab:overlapsRRCoulomb} Overlaps and energy variances calculated using Read-Rezayi state ($\ket{\psi_{r}}$), obtained for the two-body $H_O$ that is a minimal deformation of the Coulomb interaction (actual optimal pseudopotentials are listed in Table IX
in Supplementary Materials). All the energy variances ($\sigma^{r~2}_E$) are given in units of $10^{-7}$.
}
\begin{center}
\resizebox{\columnwidth}{!}{%
\begin{tabular}{|c|c|c|c|c|c|c|c|}
\hline
$N^{max}_{h~tr}$ 							& \multicolumn{2}{c|}{8 } 					& \multicolumn{2}{c|}{10 }			& \multicolumn{2}{c|}{12}	\\

\hline
$N_{h~test}$								& $\braket{\psi_{o}|\psi_{r}}$ & $\sigma^{r~2}_E$  	& $\braket{\psi_{o}|\psi_{r}}$&$\sigma^{r~2}_E$	& $\braket{\psi_{o}|\psi_{r}}$&$\sigma^{r~2}_E$ \\
\hline
4								& 0.99978& 1.4						& 0.99752& 1.8							& 0.99956 & 0.16\\
6								& 0.98361& 10						& 0.96482& 7.1 						& 0.98541 & 1.6 \\
8								& 0.99808& 3.4 					& 0.97789& 2.4							& 0.98831 & 1.9 \\ \cline{2-3}
10								& 0.79081& 19						& 0.98560& 2.5 						& 0.95757 & 1.9 \\  \cline{4-5}
12								& 0& 20							& 0.96330& 3.1							& 0.97521 & 1.6 \\ \cline{6-7}
14								& 0.37100& 18						& 0.89771& 3.1							& 0.94460 & 1.6 \\ 

\hline
\end{tabular}
}
\end{center}
\end{table}

Overall, we find high overlaps on the test systems beyond those used to learn the generating model and conclude that the framework can be used to construct a good approximation of a wavefunction at higher system size by learning the generating Hamiltonian on a smaller and diagonalizing it for a larger system. An important prerequisite for this is that the training system is large enough to have developed all the correlations characteristic of the underlying phase of matter.

\subsection{Optimal experiment design \label{sec:wk}}

Suppose we have a quantum system, for which we can control a few (experimental) parameters, and a theoretical model that translates these parameters into a Hamiltonian. How do we choose the values of the parameters such that the system has the best possible value of a certain observable of interest?

As an example, let's consider FQHE at filling factor $\nu=12/5$. The two experimental parameters will be the finite width of the quantum well in which the quasi-2D electron gas resides, parametrised by $w$ - the width of an equivalent infinitely deep quantum well (details on such parametrisation can be found in the appendix of Ref.~\cite{fivehalfPRXPakrouski2015}); and the strength of Landau level mixing (LLM) $\kappa$ that is directly controlled by the strength of the applied magnetic field: $\kappa \sim \frac{1}{\sqrt{B}}$.

As the model translating these two experimental parameters into a Hamiltonian we will use the continuous version of the LLM model derived in Refs.~\cite{SimonRezayiMixingCorrections2013,SodemannMacDonaldMixingCorrections2013,PetersonNayakMixingCorrections2013}. In particular, we use the pseudopotential corrections presented in Ref.~\cite{PetersonNayakMixingCorrections2013} for the five values of the finite width ($w=0,1,2,3,4$) to interpolate the pseudopotentials corrections by a 4th order polynomial (see Table X
in Supplementary Materials for details).
In a nutshell, the resulting continuous model provides the values for the 2- and 3-body pseudopotentials as a function of the effective width and LLM strength: $V_m=V_m(w,\kappa)$. Now, instead of varying the pseudopotentials as in the previous sections we will vary the effective width $w$ and LLM strength $\kappa$: $H=H(w,\kappa)$.

The loss function combines the extrapolated (to the infinite system size) neutral gap (weight -100), the overlap with the model non-Abelian particle-hole conjugate of the Read-Rezayi state (weight -3.7) and the total angular momentum quantum number $L$ measured in the ground state (weight 2). So the question we are asking is: "What are the experimental parameters $w,\kappa$ most suitable for observing the most stable (high neutral gap) non-Abelian physics (high overlap with RR)?" Including $L$ we ensure that we are looking for spacially uniform states with $L=0$, which is the basic property of all valid FQHE states.

 \begin{figure}[h!]
                \centering
                \iftoggle{localCompile}{
                \includegraphics[width=9cm]{pix/wkExtrGapHeatMap}
                }{
                 \includegraphics[width=9cm]{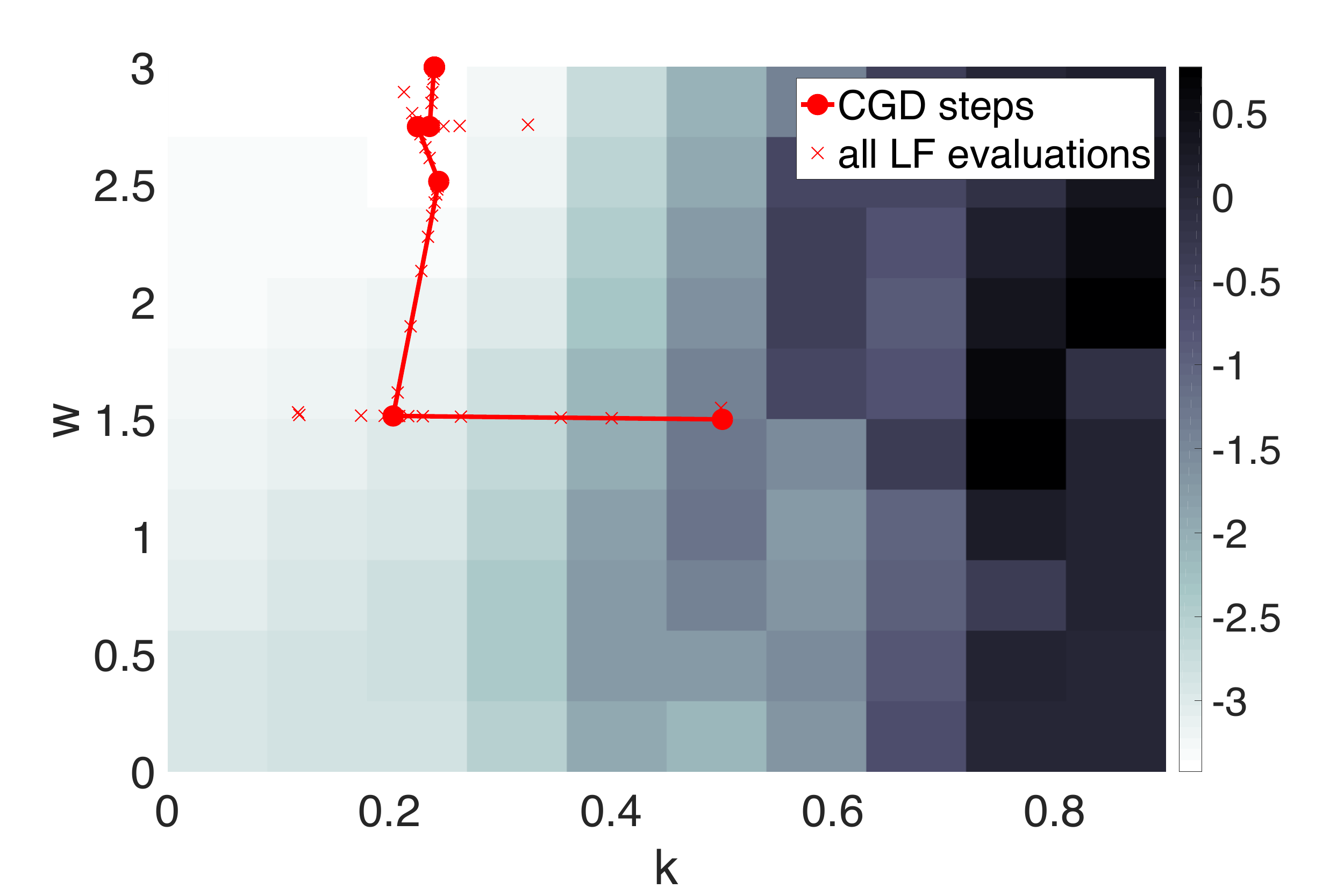}
                }
                \caption{Landscape defined by the loss function in the plane of finite width ($w$) and Landau level mixing strength ($\kappa$). Solid red dots show the history of the CGD updates from the starting point in the middle $(\kappa,w)=(1.5,0.5)$ to the optimal value $(\kappa,w)=(0.240203,2.999999)$. Red crosses indicate all the positions (118) where the loss function was evaluated to estimate the gradient and to decide on the step size.}
                \label{fig:wkHeatMap}
\end{figure}  

To visualise the landscape we search we first calculate the required observables (for the system sizes with 4,6,8 and 10 electrons) on a 2D $10\times10$ grid. The optimisation landscape defined by the loss function is shown in colour in Fig.~\ref{fig:wkHeatMap}, where we may guess a minimum around $(w,\kappa)=(3,0.2)$. The plots of each individual observable ($L$, extrapolated gap and model state overlap can be found in Fig. 1
in the Supplementary Materials.

The overlap (the loss function) has a maximum (minimum) along a narrow stripe around $\kappa\sim0.2$. This creates a long valley, a generic situation which originally motivated the development of the CGD algorithm. Taking into account the value of the gradient on the previous step it is able to go along the valley reaching the optimal point on the 5th step. A simple non-linear gradient descent on the other hand keeps bouncing between the sides of the valley (see the Fig.~2 in Supplementary Materials) and doesn't reach the optimal point after making 24 steps.

In this 2-parameter example the number of loss function evaluations (118) during optimisation was comparable to the grid search (100). However the optimal point was found here to machine precision whereas it would only be determined up to the grid step for the grid search. The real advantage comes in situations when more parameters need to be optimised, where the conjugate gradient descent will scale linearly (gradient estimation) with the parameter space dimension as opposed to the exponential scaling of the grid search.

Note that in this example we have limited the range of the parameters $\kappa$ and $w$ to the region where the model is expected to be valid: [0,1]$\times$[0,3]  by setting the value of the loss function to predefined unfavourable values which encourages the optimisation to stay within the chosen region. In the same manner in general setting the parameter values may be limited such that we find the best solution in the experimentally accessible region of the parameter space.

The results of this calculation can be directly used as a guidance for experiment: within the realistic model used~~\cite{PetersonNayakMixingCorrections2013} the best conditions for observing non-Abelian physics of the $\nu=12/5$ FQHE in GaAs samples are achieved for the effective width of the quantum well $w=3$ and LLM parameter $\kappa=0.24$.

\section{Future applications}
We've presented a general optimisation framework for searching for optimal Hamiltonians given a variational ansatz for the Hamiltonian and a replaceable numerical tool (back-end) that is able to extract the quantity of interest from it. Combining multiple properties in the loss function allows for virtually infinite number of possible applications four of which: finding parent Hamiltonians for model wavefunctions; search for simpler, experimentally accessible Hamiltonians implementing given phase of matter; extrapolating model wavefunctions to higher system sizes and finding optimal experimental parameters, are discussed here.

One may reduce the search space introducing a constraint as a dominant term in the loss function. For example, one could start from a Hamiltonian that is known to be gapped and topological and add gap to the loss function with a large negative weight. As a result one will be searching within the subspace adiabatically connected to the starting Hamiltonian. Similarly, adding a large term proportional to the properly defined distance to a given phase one will be searching within the specific phase on the phase diagram. This may result in a discovery of new model Hamiltonians as special points within the search subspace.

Quantum simulation experiments with control over several interaction terms may profit by identifying the optimal way of combining these terms in order to obtain the desired many-body phase or maximise given observables.

To infer the Hamiltonian implemented in quantum hardware one could perform a number of measurements on a system and make the loss function proportional to the total of absolute values of deviations between the actually measured value and the value given by the variational Hamiltonian thereby fitting a Hamiltonian ansatz to reproduce the measurements.

The performance and scaling of the method may be improved by deriving analytical expressions for the partial derivatives of the loss function with respect to the variational parameters. An alternative approach is to use a numerical back-end supporting automatic differentiation~\cite{AutomaticDifferentiationInML,LeiDifferProgrTensNetw}.

\section{Acknowledgements}
At the initial stage this work was supported by a grant from the Swiss National Supercomputing Centre (CSCS) under Project IDs s395 and s551.
The simulations presented in this article were performed on computational resources managed and supported by Princeton's Institute for Computational Science $\&$ Engineering and OIT Research Computing.
KP was supported by the Swiss National Science Foundation through the Early Postdoc.Mobility grant P2EZP2$\_$172168. This work was also supported in part by DOE grant No. DE-SC0002140.

\iftoggle{pasteBBLHere}{

\include{plaintexBBL.tex}

}{
\bibliographystyle{plainnat}
\bibliography{methodArxivV2}

}

\end{document}